\documentclass[11pt, a4paper]{article}
\usepackage{graphicx}%
\usepackage{multirow}%
\usepackage[authoryear,round]{natbib}
\usepackage{amsmath,amssymb,amsfonts}%
\usepackage{amsthm}%
\usepackage{mathrsfs}%
\usepackage[title]{appendix}%
\usepackage{xcolor}%
\usepackage{textcomp}%
\usepackage{manyfoot}%
\usepackage{booktabs}%
\usepackage{algorithm}%
\usepackage{algorithmicx}%
\usepackage{algpseudocode}%
\usepackage{listings}%
\usepackage{hyperref}%
\usepackage{float}%
\usepackage{subcaption}
\usepackage{authblk}
\usepackage{threeparttable}

\begin{document}

\title{Relativistic accretion-disc models of Cyg X-1\\ in the soft state: a case study} 

\author[1]{Luciano Ye}
\author[2]{Luigi Foschini}

\affil[1]{Scuola Universitaria Superiore IUSS Pavia, Piazza della Vittoria 15, Pavia, 27100, Italy}
\affil[2]{Osservatorio Astronomico di Brera, Istituto Nazionale di Astrofisica (INAF), Via E. Bianchi 46, Merate, 23007, Italy}

\date{\today}

\maketitle

\begin{abstract}
    The thermal continuum of Cygnus X-1 in the soft state is often modelled with a 
    multicolour disc in a Newtonian spacetime approximation. That choice is convenient 
    for estimating the inner radius, but it neglects relativistic effects near the 
    black hole; these are usually restored through colour-correction and 
    boundary-condition factors. Using \emph{Suzaku} data from one exemplary observation, 
    we compare this Newtonian description with some fully relativistic disc 
    models, keeping the same treatment of absorption, ionised winds, Comptonization 
    and reflection. With the mass, distance and inclination fixed to the best 
    current dynamical values, the Newtonian inner radius is consistent with a 
    high-spin Kerr innermost stable circular orbit. The relativistic models do not provide a 
    uniformly better description of the spectrum: some worsen the fit or leave spin, 
    inclination and hardening unconstrained, while the formally best model requires 
    an inclination much larger than the dynamical one or a smaller spin. 
    Apparent tensions with the models arise from data 
    degeneracies and from the limited energy band, not from a failure of general 
    relativity. 
\end{abstract}

\section{Introduction}
General relativity is one of the most tested theories. In particular, 
in recent decades, efforts have increased to find hints for developing 
a theory for the quantisation of gravity. One possible approach is to test the
spacetime around stellar-mass black holes, where the curvature is stronger.
\citet{FOSCHINI2024} studied a sample of 
stellar-mass black holes, measured the radius of the innermost stable circular orbit 
$r_{\rm ISCO}$, and compared it with the predictions of general relativity. 
They calculated $r_{\rm ISCO}$ using the multicolour blackbody of the accretion 
disc (\texttt{diskbb} model in \texttt{XSPEC}, \citealt{MITSUDA1984}), which is a 
Newtonian approximation. 
After applying the required corrections, namely the hardening factor 
\citep{SHIMURA1995} and the appropriate boundary condition correction \citep{KUBOTA1998}, 
they found no significant deviations
from the expectations of general relativity.  
Even though the \texttt{diskbb} model is well known, reliable, and provides
quite a good approximation, the natural next step is to use
fully relativistic models to understand whether there are any significant improvements or 
deviations. This is the aim of the present work. 

Therefore, we decided to reanalyse the \emph{Suzaku} observation of Cyg X-1 in the soft state
described by \citet{TOMSICK2014} and identified by \citet{FOSCHINI2024}
as the most interesting case, because the inner disc extends down to $r_{\rm ISCO}$. 
We downloaded the publicly available data, 
reprocessed them, extracted the spectra, and fitted them with five fully relativistic 
accretion disc models. 

The reference measurements of Cyg X-1 were taken from \citet{MILLERJ2021} and 
are shown in Table \ref{tab:reference_values}.

\begin{table}[H]
\centering
\caption{
        Reference values of Cyg X-1 provided by \citet{MILLERJ2021}.
        $i$ is the inclination of the accretion disc, $d$ is the distance, 
        $M$ is the mass of Cyg X-1, $a$ is the spin,
        $R_{\rm g}$ is the gravitational radius and
        $r_{\rm ISCO}$ is the radius of the innermost stable circular orbit
        according to the Kerr metric \citep{BARDEEN1972}.
        }
\label{tab:reference_values}
\Large
\begin{tabular}{l l l}
\toprule
\textbf{Parameter} & \textbf{Value} & \textbf{Unit} \\
\midrule
$i$ & $27.51^{+0.77}_{-0.57}$ & deg \\
$d$ & $2.22_{-0.17}^{+0.18}$ & kpc \\
$M$ & $21.2 \pm 2.2$ & $M_\odot$ \\
$a$ & 0.9696 -- 0.9985 & \\
$R_{\rm g}$ & $31 \pm 3$ & km \\
$r_{\rm ISCO}$ & 1.21 -- 1.74& $R_{\rm g}$ \\
\bottomrule
\end{tabular}
\medskip
\footnotesize
\end{table}

\section{Observation and data reduction}

We retrieved the \emph{Suzaku} observation of Cyg X-1 
performed on 31 October 2012 (ObsID 407072010, \citealt{TOMSICK2014}) from the HEASARC public archive.
XIS0 and XIS1 were operating in 1/4-window mode, 
with a readout frame of 2 s per CCD, while the effective exposure per frame was 
limited to 0.135 s using the burst option, yielding a total effective exposure time of $\approx 2$~ks.
Meanwhile, XIS3 was configured in burst-clock mode without the window option.
Each frame had a readout time of 8 s and an effective exposure time of only 0.1 s (exposure time $\approx 0.37$~ks), 
with a delay of 7.9 s (\citet{TOMSICK2014} did not 
analyse XIS3 data because they believed it was operating in continuous readout mode).
XIS2 had not been operational since 2006. We reprocessed the XIS data using HEASoft version 
6.36 with the calibration files updated on 2018-10-23 and following the standard procedure 
described in the \href{https://heasarc.gsfc.nasa.gov/docs/suzaku/analysis/abc/}{\emph{Suzaku} Data Reduction Guide}.

Given the high flux, on the order of $10^{-8}$~erg~cm$^{-2}$~s$^{-1}$, the XIS 
detectors were affected by pile-up.
Therefore, we extracted the source counts from an annular region with an
inner radius of 105 arcseconds following the prescription of \cite{YAMADA2012}.
The outer radius was selected depending on the instrument setting: in the case of XIS0 
and XIS1, which were operated in 1/4-window mode, 
the outer radius was 120 arcseconds due to the window boundaries; 
in the case of XIS3, where the full detector 
was available, the outer radius was 260 arcseconds.
The background was extracted from a rectangular region in a source-free zone of the 
detectors. The spectra were grouped to have at least 
30 counts per bin. We added a 2.5\% systematic error to use $\chi^2$ statistics
and fitted the spectra in the 1.2--10.0 keV energy band, to take into account pile-up residuals.
We noted that Tomsick et al. removed the 1.7--2.1 keV band because of calibration 
uncertainties and the band above 9 keV because of pile-up residuals, likely
due to a smaller extraction region. In our case, we were able to recover these energy 
bands, because we had better calibration (2018 vs 2014).

All fitted parameters in this article have 90\% confidence intervals.

\section{Baseline models}
The first step is to reproduce the spectral model proposed by \citet{TOMSICK2014}, 
who used \texttt{diskbb} (\citealt{MITSUDA1984} and \citealt{MAKISHIMA1986}) 
plus a physical model of Comptonization.
Tomsick et al. performed a joint spectral fit using \emph{Suzaku} and \emph{NuSTAR} data. 
We noted (see Figure 3 of \citealt{TOMSICK2014}) that 
the statistics of \emph{NuSTAR} are much better than those of \emph{Suzaku}, 
resulting in better-defined iron lines. 
In addition, the \emph{NuSTAR} spectra extend up to about 80 keV, making it easier to 
define the Comptonized continuum.
In our case, since we use only \emph{Suzaku} data, 
we decided to use phenomenological models for Comptonization. 
We found a good fit with the following model, which we consider to be the \textbf{baseline model}: 

\begin{equation*}
\texttt{constant * TBabs * zxipcf * (powerlaw + reflionx + diskbb)}
\end{equation*}

\begin{figure}[H]
    \centering
    \includegraphics[width=1\textwidth]{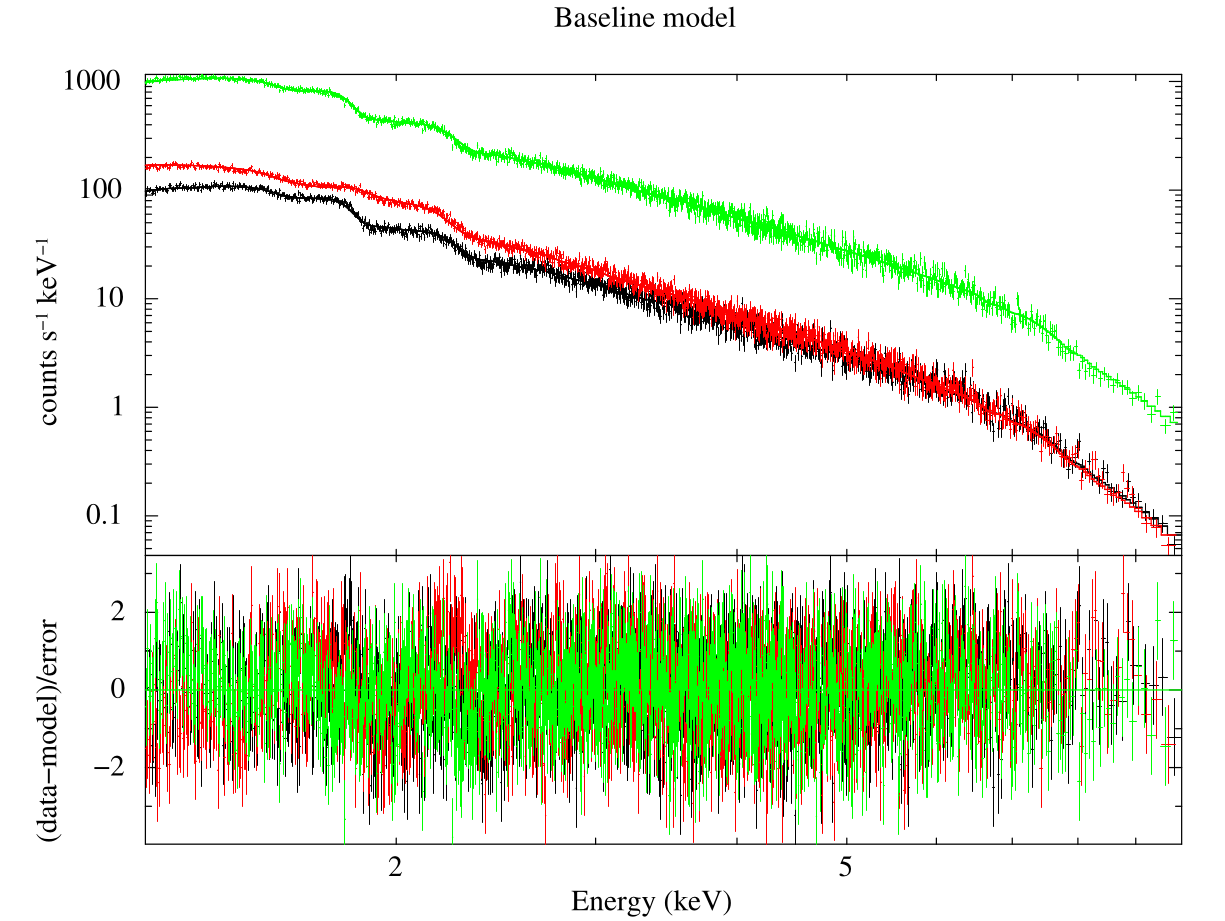}
    \caption{\emph{Suzaku} XIS spectra fitted with the baseline model. 
    The top panel shows the data and the best-fit model, 
    while the bottom panel shows the residuals in units of $\sigma$.}
    \label{fig:diskbb}
\end{figure}

For the Galactic absorption (\texttt{TBabs, \citealt{WILMS2000}}), 
we used the value of $7.05 \times 10^{21}$~cm$^{-2}$ \citep{HI4PI2016}.
The Comptonization, the reflection, and the absorption by ionised winds were modelled by 
\texttt{powerlaw}, \texttt{reflionx} (\citealt{ROSS1999}, \citealt{ROSS2005}) 
and \texttt{zxipcf} (\citealt{MILLER2006} \citealt{REEVES2008}).
We then tied the photon index $\Gamma$ of \texttt{reflionx} to that of \texttt{powerlaw},
and the covering fraction of \texttt{zxipcf} was set to 1 to stabilise the fit,
because we noted that it always reached the upper limit.

The result of this model is shown in Table \ref{tab:diskbb_fit}.
The disc peak temperature is \(0.50 \pm 0.01\) keV and the normalisation is $3.56^{+0.22}_{-0.20} \times 10^4$.
Since Cyg X-1 is in the soft state, the inner radius $r_{\rm in}$ is
assumed to be equal to the innermost stable circular orbit $r_{\rm ISCO}$ and can be
calculated with the following formula \citep{KUBOTA1998}:

\begin{equation}
    \label{eq:diskbb_norm}
    r_{\rm in} = \kappa f^2 d_{10} \sqrt{\frac{N}{\cos i}}
\end{equation}

where $\kappa = 0.412$ is the boundary-condition factor \citep{KUBOTA1998}, $f=1.7$ is 
the spectral hardening factor \citep{SHIMURA1995}, 
$N$ is the normalisation of the \texttt{diskbb} model, $i$ is the inclination 
and $d_{10}$ is the distance in units of 10 kpc.

By inserting the values from Table \ref{tab:reference_values} into Eq.~\ref{eq:diskbb_norm}, we obtained 
$r_{\rm in} = 53^{+5}_{-4}$~km $= 1.7 \pm 0.3 R_{\rm g}$,
whereas by using $N$ measured by \citet{TOMSICK2014}, we obtained 
$r_{\rm in} = 40 \pm 3$~km $ = 1.3 \pm 0.2 R_{\rm g}$.
The iron abundance is larger than that calculated by \cite{TOMSICK2014}, 
but it depends highly on the continuum models \citep{DURAPHE2026}.
The characteristics of the disc obtained by \texttt{diskbb} are comparable to 
those obtained by \citet{TOMSICK2014};
thus the baseline model is reliable for describing the accretion disc,
even though our Comptonization is a phenomenological model. An improper treatment of the 
Comptonization would lead to a decrease in $r_{\rm in}$, 
not an increase \citep{MERLONI2000}.

We also tried to fix the ionisation parameter $\xi$ of \texttt{reflionx} to 
that of \texttt{zxipcf}, but the fit was worse.

\begin{table}[H]
\centering
\caption{\texttt{constant * TBabs * zxipcf * (powerlaw + diskbb + reflionx)}\\
The fit yields \(\chi^2 = 3514.14\) for 3642 d.o.f.\\
$N_H$ is the hydrogen column density, $\xi$ is the ionisation parameter,
$kT_{in}$ is the temperature at the inner disc radius, $\Gamma$ is the photon index
and $N$ is the normalisation of \texttt{diskbb}.
}
\label{tab:diskbb_fit}
\begin{tabular}{l l c l}
\toprule
\textbf{Component} & \textbf{Parameter} & \textbf{Unit} & \textbf{Value} \\
\midrule
zxipcf & \(N_{\rm H}\) & \(10^{22}\,\mathrm{cm^{-2}}\) & $4.3^{+1.5}_{-1.2}$ \\
zxipcf & \(\log\xi\) & ---  & $3.56^{+0.07}_{-0.04}$\\
powerlaw & $\Gamma$ & --- & $3.11 \pm 0.02$\\
powerlaw & norm & ph s\(^{-1}\) cm\(^{-2}\) keV\(^{-1}\)  & $28.8^{+1.5}_{-1.6}$\\
diskbb & \(kT_{\rm in}\) & keV  & \(0.50 \pm 0.01\) \\
diskbb & $N$ & $10^{4}$ & $3.56^{+0.22}_{-0.20}$\\
reflionx & Fe abundance & solar &  $5.0^{+1.1}_{-0.6}$\\
reflionx & norm & $10^{-4}$ & $8^{+3}_{-4}$\\
reflionx & $\xi$ & erg cm s$^{-1}$ & $1300^{+800}_{-200}$ \\
\bottomrule
\end{tabular}
\medskip
\footnotesize
\end{table}

\section{Relativistic accretion-disc models}

After establishing the baseline model, \texttt{diskbb} was replaced with other 
fully relativistic accretion disc models. 
First of all, we replaced \texttt{diskbb} with \texttt{kerrd},
which is an extreme-Kerr disc model based on the transfer function explained by \citet{LAOR1991}
and directly measures the inner radius $r_{\rm in}$.
The distance, the mass and the inclination were fixed according to Table \ref{tab:reference_values}
and the hardening factor was set to the standard value 1.7.

Table \ref{tab:kerrd_fit} shows that $r_{\rm in} <2.6$~\(R_g\) (corresponding to $a > 0.86$),
which is consistent with the value calculated using the baseline model and with the 
Eddington ratio $L/L_{\rm Edd} \approx 1.8 \%$, which is a reasonable value for the soft state.
Additionally, the shift in wind parameters can be explained by the results of \cite{DURAPHE2026}.
However, the fit seems to be worse, because
$\chi^2$ increases by approximately 110 for the same number of d.o.f.

\begin{table}[H]
\centering
\caption{\texttt{constant * TBabs * zxipcf * (kerrd + powerlaw + reflionx)}\\
\(\chi^2 = 3624.16\) for 3642 d.o.f.,
$\dot{M}$ is the accretion rate and $r_{\rm in}$ is the inner radius.\\
The normalisation and the outer radius of \texttt{kerrd} are fixed to 1 and $10^5$~$R_{\rm g}$}
\label{tab:kerrd_fit}
\begin{tabular}{l l c l}
\toprule
\textbf{Component} & \textbf{Parameter} & \textbf{Unit} & \textbf{Fit 1}\\
\midrule
zxipcf & \(N_{\rm H}\) & \(10^{22}\,\mathrm{cm^{-2}}\)  & $4.5 \pm 0.6$\\
zxipcf & \(\log\xi\) & --- &  $3.15^{+0.5}_{-0.7}$\\
kerrd & \(\dot{M}\) & \(10^{18}\,\mathrm{g\,s^{-1}}\)  & $0.194^{+0.007}_{-0.006}$ \\
kerrd & $r_{\rm in}$ & \(R_{\rm g}\)  & $<2.6$\\
powerlaw & $\Gamma$ & --- &  $3.09^{+0.02}_{-0.01}$ \\
powerlaw & norm & ph s\(^{-1}\) cm\(^{-2}\) keV\(^{-1}\) & $31.6^{+2.0}_{-0.8}$\\
reflionx & Fe abundance & solar &  $2^{+2}_{-1}$\\
reflionx & Norm & $10^{-4}$ &  $120^{+250}_{-70}$ \\
reflionx & $\xi$ & erg cm s$^{-1}$ &  $30^{+30}_{-20}$ \\
\bottomrule
\end{tabular}
\medskip
\footnotesize
\end{table}

We also tried to let the inclination vary, since \citet{TOMSICK2014} obtained
different values.
As a result, the fit did not converge, and the error intervals of $\dot{M}$ and $r_{\rm in}$
were not calculated, although the fit ended with $r_{\rm in} \approx 5.5 R_{\rm g}$, 
$i = 51^{+8}_{-1}$~deg, $\dot{M}=0.28 \times 10^{18}$~g~s$^{-1}$ 
and $\Delta \chi^2 = - 7.18$ with 1 fewer d.o.f. compared with the baseline model.
Then, we allowed the hardening factor $f$ to vary,
but the error intervals of $r_{\rm in} = 5.5$~$R_{\rm g}$ could not be calculated.

Turning to the \texttt{kerrbb} model \citep{KERRBB}, we again fixed the distance, the mass
and the spectral hardening factor, and we assumed zero torque at the inner boundary,
so the $\eta$ parameter was set to 0.
However, Fit 2 (Table \ref{tab:kerrbb_fit}) showed that $\dot{M}$ and $a$ 
were fitted with reasonable values ($L/L_{\rm Edd} = 1.6\%$), 
but their error intervals were not calculated and $\chi^2$ increased by 117.38 
with the same number of d.o.f.
Freeing the inclination, we obtained a better fit (Fit 3, $\Delta \chi^2 = -7.06$ with 2 fewer d.o.f. 
compared with the baseline model), but $\dot{M}$ and $a$ were still unconstrained although
the values were reasonable.

\begin{table}[H]
\centering
\caption{\texttt{constant * TBabs * zxipcf * (powerlaw + kerrbb + reflionx)}\\
Fit 2: \(\chi^2 = 3631.52\) for 3642 d.o.f.
Fit 3: \(\chi^2 = 3507.08\) for 3640 d.o.f., $a$ is the spin,
$r_{\rm ISCO}$ is the inner radius calculated according to the Kerr metric.}
\label{tab:kerrbb_fit}
\begin{threeparttable}

\begin{tabular}{l l c l l}
\toprule
\textbf{Component} & \textbf{Parameter} & \textbf{Unit} & \textbf{Fit 2} & \textbf{Fit 3}\\
\midrule
zxipcf & \(N_{\rm H}\) & $10^{22}$~cm$^{-2}$ & $1.1^{+0.1}_{-0.2}$ & $4.7^{+1.8}_{-1.5}$\\ 
zxipcf & \(\log\xi\) & --- & $2.2 \pm 0.1$ & $3.56^{+0.8}_{-0.5}$\\
powerlaw & $\Gamma$ & --- & $3.2 \pm 0.3$ & $3.05^{+0.1}_{-0.3}$\\
powerlaw & norm & ph s\(^{-1}\) cm\(^{-2}\) keV\(^{-1}\) & $36^{+1}_{-2}$ & $25 \pm 2$\\
kerrbb & Spin \(a\) & --- & 0.9999\tnote{a} & 0.9999\tnote{a}\\
kerrbb & \(\dot{M}\) & \(10^{18}\,\mathrm{g\,s^{-1}}\) & 0.13\tnote{a} & 0.09\tnote{a}\\
kerrbb & $i$ & deg & 27.51 \tnote{b} & $40.5^{+0.7}_{-1.6}$\\
kerrbb & Norm & --- & 1 \tnote{b} & $1.3^{+0.2}_{-0.1}$\\
reflionx & Fe abundance & solar &  $2.4^{+1.2}_{-0.5}$ & $5.0^{+1.5}_{-0.4}$\\
reflionx & Norm & $10^{-4}$ & $4.8^{+2.8}_{-0.8}$ & $9.4^{+2.6}_{-2.4}$\\
reflionx & $\xi$ & erg cm s$^{-1}$ & 2000\tnote{a} & $1100 \pm 100$ \\
--- & $r_{\rm ISCO}$ & $R_{\rm g}$ &  1.08 & 1.08 \\
\bottomrule
\end{tabular}
\begin{tablenotes}
\item[a] pegged
\item[b] frozen
\end{tablenotes}
\medskip
\footnotesize
\end{threeparttable}
\end{table}

Then we used the \texttt{bhspec} model \citep{DAVIS2005}, where we set the value of $\log(M/M_\odot)$
and the normalisation according to Table \ref{tab:reference_values}.
At first, we tried to freeze $\cos i$ to 0.89, but the fit ended with
the normalisation and iron abundance of \texttt{reflionx} tending to 0.

Thus, we freed the inclination (Fit 4) and the result was much better than that of the baseline model
with a change in $\Delta \chi^2 = -29.15$ and a reduction of 1 d.o.f. (see Table \ref{tab:bhspec_fit}).
Moreover, $L/L_{\rm Edd} = 2.7\%$, but the spin and the inclination were not consistent
with those of \citet{TOMSICK2014}, \citet{ZDZ2024A} and \citet{ZDZ2024B}.
We then tried to freeze the spin (Fit 5) to the average measured value \citep{MILLERJ2021}
and obtained $i = 49^\circ$.
The fit was still good and comparable to the baseline model, 
with a decrease in $\chi^2$ of 2.72 for the same number of d.o.f.
and $L/L_{\rm Edd} = 1.3 \%$.

\begin{table}[H]
\centering
\caption{\texttt{constant * TBabs * zxipcf * (powerlaw + bhspec + reflionx)}\\
Fit 4: \(\chi^2 = 3484.99\) for 3641 d.o.f.,
Fit 5: \(\chi^2 = 3511.42\) for 3642 d.o.f.,
$L/L_{\rm Edd}$ is the Eddington luminosity ratio,
$r_{\rm ISCO}$ is the inner radius calculated according to the Kerr metric.
}
\label{tab:bhspec_fit}
\begin{threeparttable}
\centering
\begin{tabular}{l l l l l}
\toprule
\textbf{Component} & \textbf{Parameter} & \textbf{Unit} & \textbf{Fit 4} & \textbf{Fit 5}\\
\midrule
zxipcf & \(N_{\rm H}\) & \(10^{22}\,\mathrm{cm^{-2}}\) & $3.8^{+1.8}_{-1.1}$ & $3.6^{+1.5}_{-0.9}$\\
zxipcf & \(\log\xi\) & --- & $3.56^{+0.10}_{-0.05}$ & $3.56^{+0.07}_{-0.04}$\\
powerlaw & $\Gamma$ & --- &$3.14 \pm 0.02$ & $3.13^{+0.02}_{-0.01}$\\
powerlaw & norm & ph s\(^{-1}\) cm\(^{-2}\) keV\(^{-1}\) & $32.8 \pm 1.6$ & $31.0^{+1.8}_{-0.5}$\\
bhspec & \(\log(L/L_{\rm Edd})\) & --- & -1.57\tnote{b} & $-1.90 \pm 0.02$\\
bhspec & \(\cos i\) & --- & $0.20^{+0.05}_{-0.01}$ & $0.65^{+0.02}_{-0.01}$\\
bhspec & spin \(a\) & --- & $0.60^{+0.04}_{-0.14}$ & 0.98\tnote{a}\\
reflionx & Fe abundance & solar & $4.5^{+1.4}_{-1.9}$ & $4.1^{+1.1}_{-0.9}$\\
reflionx & norm & $10^{-4}$  & $3.0^{+0.8}_{-1.0}$ & $2.9^{+0.4}_{-1.0}$\\
reflionx & $\xi$ & erg cm s$^{-1}$ & $2100^{+200}_{-1000}$ & $2100 \pm 100$ \\
--- & $r_{\rm ISCO}$  & $R_g$ & \(3.83^{+0.56}_{-0.17}\) &\(1.61\) \\
\bottomrule
\end{tabular}
\begin{tablenotes}
    \item[a] frozen
    \item[b] pegged
\end{tablenotes}
\medskip
\footnotesize
\end{threeparttable}
\end{table}

We also tried the \texttt{slimdisk} model, created by \citet{KAWAGUCHI2003}, but it is clear
that it is not suited to the spectra, since it requires a very high accretion rate.
All seven fits yield unrealistic physical parameters.

In the case of \texttt{superkerr} \citep{MUMMERY2024}, 
the ISCO stress parameter $\delta_j$,
which is the fraction of angular momentum the fluid passes back to the disc,
was initially set to 0, meaning that there is no communication between the disc and the plunging region 
and that the ISCO stress vanishes.
In this case, too, the fit returned unphysical parameters.

\section{Discussion}

Using \emph{Suzaku} XIS data, we have presented fits to the spectrum with five fully relativistic models, 
replacing \texttt{diskbb} in the baseline model while keeping 
the same treatment of Comptonization, reflection, absorption, and ionised winds.

In general, the inclination, the mass and the distance are fixed to the best current
values and the Newtonian radius is comparable to a high-spin Kerr innermost stable circular orbit.
However, relativistic models do not always provide a better description of the spectrum.
For instance, we are not able to measure the value of the inclination as well 
as that of \cite{MILLERJ2021}, who obtained the value by modelling optical photometric 
measurements to study the inclination of the binary system, whereas
\texttt{kerrbb} and \texttt{bhspec} measure the inclination of the inner disc,
which is probably misaligned with the binary system. However, the 
range we obtained (40$^\circ$ -- 78$^\circ$) is consistent with 
the values of \citet{TOMSICK2014} ($42^\circ$ -- $69^\circ$).

Moreover, either the spin is not well constrained (\texttt{kerrd}: $a>0.86$, 
\texttt{kerrbb}: $a<0.9999$) or the calculated values (\texttt{bhspec}: $a=0.60$) 
are inconsistent with those of other works ($0.9696 < a < 0.9985$ \citep{TOMSICK2014},
$0.82<a<0.997$ \citep{ZDZ2024A}, $a>0.87$ \citep{ZDZ2024B} and $a>0.97$ \citep{SALVESEN2020}).

Conversely, the parameters of the non-disc components remain stable, except for the 
iron abundance and the wind parameters, which vary among the models. 
However, they depend strongly on continuum models \citep{DURAPHE2026}.

The apparent tension between the model parameters and the observations is 
likely due to the quality of the data and the limited energy band, not to a failure of 
general relativity. It should be possible to significantly improve the results 
by using higher-quality data (e.g., \emph{NuSTAR}).


\begin{thebibliography}{00}
\bibitem[Bardeen et al. (1972)]{BARDEEN1972} J. M. Bardeen et al.,
Rotating Black Holes: Locally Nonrotating Frames, Energy Extraction, and Scalar Synchrotron Radiation,
Astrophysical Journal, Vol. 178, pp. 347-370 (1972),
\href{https://doi.org/10.1086/151796}{https://doi.org/10.1086/151796}
\bibitem[Davis et al. (2005)]{DAVIS2005} S. W. Davis et al.,
Relativistic Accretion Disk Models of High-State Black Hole X-Ray Binary Spectra,
The Astrophysical Journal, Volume 621, Issue 1, pp. 372-387,
\href{https://doi.org/10.1086/427278}{https://doi.org/10.1086/427278}
\bibitem[Duraphe et al. (2026)]{DURAPHE2026} K. Duraphe et al.,
A NuSTAR Reflection-Spectroscopy Survey of Cygnus X-1,
under review at ApJ.
\href{https://doi.org/10.48550/arXiv.2608.15902}{https://doi.org/10.48550/arXiv.2608.15902}
\bibitem[Foschini, et al. (2025)]{FOSCHINI2024} L. Foschini, A. Vecchiato and A. Bonanno,
Searching for quantum-gravity footprint around stellar-mass black holes,
European Physical Journal C, vol. 85, (2025), id 752,
\href{https://doi.org/10.1140/epjc/s10052-025-14477-3}{https://doi.org/10.1140/epjc/s10052-025-14477-3}
\bibitem[HI4PI Collaboration (2016)]{HI4PI2016} HI4PI Collaboration,
HI4PI: A full-sky H I survey based on EBHIS and GASS 
Astronomy \& Astrophysics, Volume 594, id.A116, 15 pp.,
\href{https://doi.org/10.1051/0004-6361/201629178}{https://doi.org/10.1051/0004-6361/201629178} 
\bibitem[Kawaguchi (2003)]{KAWAGUCHI2003} T. Kawaguchi,
Comptonization in Super-Eddington Accretion Flow and Growth Timescale of Supermassive Black Holes,
The Astrophysical Journal, Volume 593, Issue 1, pp. 69-84,
\href{https://doi.org/10.1086/376404}{https://doi.org/10.1086/376404}
\bibitem[Kubota, et al. (1998)]{KUBOTA1998} A. Kubota et al., 
Evidence for a black hole in the X-ray transient GRS 1009-45, 
Publ. Astron. Soc. Japan 50 (1998) 667. 
\href{https://doi.org/10.1093/pasj/50.6.667}{https://doi.org/10.1093/pasj/50.6.667}
\bibitem[Laor (1991)]{LAOR1991} A. Laor,
Line Profiles from a Disk around a Rotating Black Hole,
Astrophysical Journal v.376, p.90,
\href{https://doi.org/10.1086/170257}{https://doi.org/10.1086/170257}
\bibitem[Li, et al. (2005)]{KERRBB} L.-X. Li et al., 
Multitemperature Blackbody Spectrum of a Thin Accretion Disk around a Kerr Black Hole: Model Computations and Comparison with Observations, 
Astrophys. J. Suppl. Series 157, (2005), 335 
\href{https://doi.org/10.1086/428089}{https://doi.org/10.1086/428089}
\bibitem[Makishima, et al. (1986)]{MAKISHIMA1986} K. Makishima et al., 
Simultaneous X-ray and optical observations of GX 339-4 in an X-ray high state, 
Astrophys. J. 308 (1986) 635. 
\href{https://doi.org/10.1086/164534}{https://doi.org/10.1086/164534}
\bibitem[Merloni et al. (2000)]{MERLONI2000} A. Merloni et al.,
On the interpretation of the multicolour disc model for black hole candidates,
Monthly Notices of the Royal Astronomical Society, Volume 313, Issue 1, March 2000, Pages 193-197,
\href{https://doi.org/10.1046/j.1365-8711.2000.03226.x}{https://doi.org/10.1046/j.1365-8711.2000.03226.x}
\bibitem[Miller et al. (2006)]{MILLER2006} L. Miller et al.,
Variable iron-line emission near the black hole of Markarian 766,
Astronomy and Astrophysics, Volume 453, Issue 1, July I 2006, pp.L13-L16,
\href{https://doi.org/10.1051/0004-6361:20065276}{https://doi.org/10.1051/0004-6361:20065276}
\bibitem[Miller-Jones, et al. (2021)]{MILLERJ2021} J.~C.~A. Miller-Jones et al., 
Cygnus X-1 contains a 21-solar mass black hole -- Implications for massive star winds, 
Science 371 (2021) 1046. 
\href{https://doi.org/10.1126/science.abb3363}{https://doi.org/10.1126/science.abb3363}
\bibitem[Mitsuda, et al. (1984)]{MITSUDA1984} K. Mitsuda et al., 
Energy spectra of low-mass binary X-ray sources observed from Tenma, 
Publ. Astron. Soc. Japan 36, (1984) 741.
\bibitem[Mummery et al. (2024)]{MUMMERY2024} A. Mummery et al.,
Testing theories of accretion and gravity with super-extremal Kerr disks,
Monthly Notices of the Royal Astronomical Society, Volume 527, Issue 3, pp.5956-5973,
\href{https://doi.org/10.1093/mnras/stad3532}{https://doi.org/10.1093/mnras/stad3532}
\bibitem[Reeves et al. (2008)]{REEVES2008} J. Reeves et al.,
On why the iron K-shell absorption in AGN is not a signature of the local warm/hot intergalactic medium,
Monthly Notices of the Royal Astronomical Society: Letters, Volume 385, Issue 1, pp. L108-L112,
\href{https://doi.org/10.1111/j.1745-3933.2008.00443.x}{https://doi.org/10.1111/j.1745-3933.2008.00443.x}
\bibitem[Ross et al. (1999)]{ROSS1999} R. R. Ross et al.,
X-ray reflection spectra from ionized slabs,
Monthly Notices of the Royal Astronomical Society, Volume 306, Issue 2, pp. 461-466,
\href{https://doi.org/10.1046/j.1365-8711.1999.02528.x}{https://doi.org/10.1046/j.1365-8711.1999.02528.x}
\bibitem[Ross et al. (2005)]{ROSS2005} R. R. Ross et al.,
A comprehensive range of X-ray ionized-reflection models,
Monthly Notices of the Royal Astronomical Society, Volume 358, Issue 1, pp. 211-216,
\href{https://doi.org/10.1111/j.1365-2966.2005.08797.x}{https://doi.org/10.1111/j.1365-2966.2005.08797.x}
\bibitem[Salvesen \& Miller (2020)]{SALVESEN2020} G. Salvesen \& J. M. Miller
Monthly Notices of the Royal Astronomical Society, Volume 500, Issue 3, January 2021, Pages 3640-3666,
\href{ https://doi.org/10.1093/mnras/staa3325}{ https://doi.org/10.1093/mnras/staa3325}
\bibitem[Shimura \& Takahara (1995)]{SHIMURA1995} T. Shimura \& F. Takahara, 
On the spectral hardening factor of the X-ray emission from accretion disks in black hole candidates, 
Astrophys. J. 445 (1995) 780. 
\href{https://doi.org/10.1086/175740}{https://doi.org/10.1086/175740}
\bibitem[Tomsick, et al. (2014)]{TOMSICK2014} J.~A. Tomsick et al., 
The reflection component from Cygnus X-1 in the soft state measured by NuSTAR and \emph{Suzaku}, 
Astrophys. J. 780 (2014) 78. 
\href{https://doi.org/10.1088/0004-637X/780/1/78}{https://doi.org/10.1088/0004-637X/780/1/78}
\bibitem[Wilms et al. (2000)]{WILMS2000} J. Wilms et al.,
On the Absorption of X-Rays in the Interstellar Medium,
The Astrophysical Journal, Volume 542, Issue 2, pp. 914-924,
\href{https://doi.org/10.1086/317016}{https://doi.org/10.1086/317016}
\bibitem[Yamada et al. (2012)]{YAMADA2012} S. Yamada et al., 
Data-Oriented Diagnostics of Pileup Effects on the \emph{Suzaku} XIS,
Publications of the Astronomical Society of Japan, Volume 64, Issue 3, 25 June 2012, 53, 
\href{https://doi.org/10.1093/pasj/64.3.53}{https://doi.org/10.1093/pasj/64.3.53}
\bibitem[Zdziarski et al. (2024A)]{ZDZ2024A} A.~A. Zdziarski et al., 
Black hole spin measurements in LMC~X-1 and Cyg~X-1 are highly model dependent, 
Astrophys. J. 962 (2024a) 101. 
\href{https://doi.org/10.3847/1538-4357/ad1b60}{https://doi.org/10.3847/1538-4357/ad1b60}
\bibitem[Zdziarski et al. (2024B)]{ZDZ2024B} A.~A. Zdziarski et al., 
What is the black hole spin in Cyg~X-1?, 
Astrophys. J. 967 (2024b) L9. 
\href{https://doi.org/10.3847/2041-8213/ad43ed}{https://doi.org/10.3847/2041-8213/ad43ed}
\end{thebibliography}
\end{document}